\documentclass[fleqn,twoside]{article}
\usepackage{espcrc2}
\usepackage{epsfig}

\newcommand{\ls}{a_s}

\def\simgt{\stackrel{>}{{}_\sim}}

\usepackage{graphicx}
\usepackage[figuresright]{rotating}

\newcommand{\AmS}{{\protect\the\textfont2
  A\kern-.1667em\lower.5ex\hbox{M}\kern-.125emS}}

\title{Stochastic Field evolution of disoriented chiral condensates}

\author{Lu\'{\i}s M. A. Bettencourt \address[MIT]{Center for Theoretical 
Physics, Massachusetts Institute of Technology, \\ 
        77 Massachusetts Avenue, Cambridge MA 02139, USA}%
        \thanks{I thank Krishna Rajagopal and Jim Steele for a stimulating 
collaboration. See \cite{BRS} for a full account of the 
results summarized here.},
        }
       
\begin{document}

\begin{abstract}
I present a summary of recent work \cite{BRS} where we describe 
the time-evolution of a region of disoriented chiral 
condensate via Langevin field equations for the linear 
$\sigma$ model. We analyze the model in equilibrium, paying attention
to subtracting ultraviolet divergent classical terms
and replacing them by their finite quantum counterparts.
We use results from lattice gauge theory and chiral
perturbation theory to fix nonuniversal constants.
The result is a ultraviolet cutoff independent theory 
that reproduces quantitatively the expected equilibrium
behavior of pion and $\sigma$ quantum fields.
We also estimate the viscosity $\eta(T)$, 
which controls the dynamical timescale in the Langevin equation, 
so that the near equilibrium dynamical response agrees with 
theoretical expectations.
\vspace{1pc}
\end{abstract}

\maketitle

The relativistic heavy ion collider (RHIC) is currently 
smashing nuclei together to produce the most extreme states 
of nuclear matter ever achieved in a controlled environment. 
Their cooling dynamics may then reveal collective properties of hot 
nuclear matter {\it viz.} its equation of state and the nature 
of the confining and chiral symmetry breaking transitions. 

Extracting this information requires dynamical extrapolation from 
the final products that hit particle detectors, across possible critical 
points, up to the initial hot phase.
This is practically impossible in general but the theory of 
critical dynamics, a byproduct of the renormalization group, 
tells us how it may be achievable by modeling the dynamics of light 
degrees of freedom as stochastic (classical) fields. In nuclear 
Physics these are the pion and $\sigma$ fields and the associated 
equations of motion are stochastic versions of the Euler-Lagrange equations 
for the linear $\sigma$-model.

The simplest Langevin prescription is to add 
stochastic sources $\xi_a(x,t)$ and dissipation terms to the classical
field equations for $\phi_a=(\sigma,{\vec \pi})$
\begin{eqnarray}
&& \frac{\partial^2\phi_a}{\partial t^2} - \nabla^2 \phi_a
+ \lambda \left( \phi_a^2 - v^2 \right) \phi_a - H \delta_{a0} \\
&& \qquad \qquad \qquad \qquad \qquad  
= -\eta_{ab} \frac{\partial \phi_b}{\partial t}  + \xi_a \ . \nonumber 
\label{langevin}
\end{eqnarray} 
The stochastic fields $\xi_a$ are taken to obey the fluctuation-dissipation 
relations
\begin{equation}\label{fluctuationdissipation}
\langle \xi_a(x) \xi_b(y) \rangle = 2 \eta_{ab}(x) T 
\;\delta^4(x-y) \ ,
\end{equation}
with $\langle \xi_a(x) \rangle =0$. Here, we shall always
choose $\eta_{ab}(x)=\eta(T)\delta_{ab}$.
This dynamics reduces to the classical microcanonical relativistic 
evolution for the pions and sigma as $\eta\rightarrow 0$.

In the limit of vanishing masses the long wavelength modes of the fields
are effectively overdamped, and suffer critical slowing down~\cite{Bett}. 
Then our equations correspond to Model A, in the classification of 
dynamical stochastic theories of Hohenberg and Halperin~\cite{HoHa}. 
Although Model A will be adequate for our purposes, a complete analysis
of the universal $O(4)$ dynamics in the chiral limit of QCD
requires coupling the order parameter (which is not conserved) to other 
conserved quantities in the theory, as in Hohenberg and Halperin's 
Model G~\cite{RajagopalWilczekStatic}.

The long time Langevin dynamics takes the system to its (classical) 
canonical ensemble, at temperature $T$. This results in divergences 
of certain field expectation values as the lattice spacing $\ls$ 
is taken to zero.
To make physical sense of the model in this limit requires 
a $T$ dependent renormalization scheme. This is achieved by making 
the replacement in the bare Eq.~(1),
\begin{eqnarray}
-\lambda v^2 \rightarrow 
-\lambda v^2 + \Delta I_{\rm tadpole} + \Delta I_{\rm sunset},
\label{counterterms}
\end{eqnarray} where
\begin{eqnarray}
&& \Delta I_{\rm tadpole} \equiv  I_{\rm tadpole}^{\rm B-E} 
- I_{\rm tadpole}^{\rm cl}, \nonumber \\
&& I_{\rm tadpole}^{\rm cl} = \frac{N+2}{2\pi^2} \lambda T \Lambda
\;\to\; 0.25\, \frac{(N+2)\lambda T}{\ls} \ , \nonumber 
\\
&& I_{\rm tadpole}^{\rm B-E} = \frac{N+2}{12} {\lambda T^2 \over \hbar}
, \nonumber \\
&& \Delta I_{\rm sunset} \to \left[0.014\, \ln ({T \ls \over \hbar}) 
- 0.037\right]
  (N+2) \lambda^2 T^2.  \qquad  \nonumber 
\end{eqnarray}
This removes linear and logarithmic divergences in the self energy of pion
and $\sigma$ fields.

\vspace{-0.9 cm}
\begin{figure}[htb]
\psfig{file=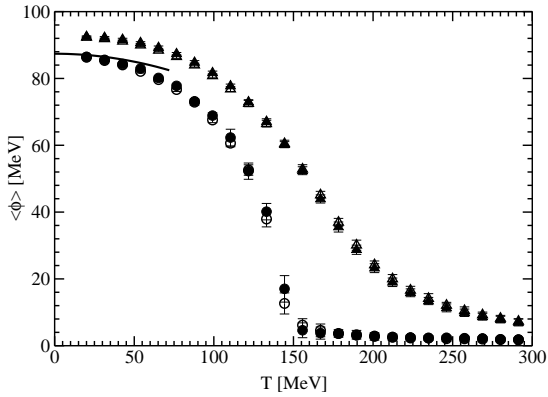,width=2.5in,angle=270}
\vspace{-1 cm}
\caption{The order parameter $\langle \phi \rangle$ {\it vs.} 
temperature $T$, for $H=0$ (circles)
and $H$ nonzero (triangles). Open and
filled symbols show results obtained with $a_s=1$ on a $26^3$ lattice
and $a_s=0.4$ on a $64^3$ lattice respectively, with counterterms
as described in the text. The line shows the prediction from chiral 
perturbation theory, Eq. (\ref{chiral}).}
\label{fig:op}
\vspace{-0.6 cm}
\end{figure}

In Fig.~\ref{fig:op}, we required that the order parameter 
$\langle \phi \rangle$ agrees
with the expectations from chiral perturbation theory at low 
temperature~\cite{chpt},
\begin{equation}
\langle \phi \rangle = f_\pi \left( 1 - \frac{T^2}{12 f_\pi^2} \right)\ ,
\qquad T\ll f_\pi \ .
\label{chiral}
\end{equation}
In particular, we have enforced the absence of linear $T$-dependence
at small $T$ by the addition of a further mass counterterm.
This counterterm is not divergent as $\ls\rightarrow 0$, but it does
vary with $\ls$. We find that the linear $T$-dependence
at small $T$ is removed by 
$-\lambda v^2 \rightarrow -\lambda v^2 + b_1 T$, with 
$b_1=0.425$ in dimensionless units.  Next, we fixed the finite
counterterm $\propto T^2$.  We note that in the absence of any finite
counterterm proportional to $T^2$ the
second order phase transition (for $H=0$) 
occurs at $T_c\simeq 130$ MeV, which is somewhat lower
than that expected in QCD~\cite{LatticeScreeningLengths,LatticeReviews}. 
We have pushed $T_c$ up to  $T_c\simeq 150$ MeV
by introducing 
$-\lambda v^2 \rightarrow -\lambda v^2 + b_1 T+ b_2 T^2$, 
with $b_2=-0.066$ in dimensionless 
units. These values of $b_1$ and $b_2$ were obtained with
a lattice spacing $\ls=0.4$ on a $64^3$ lattice, as shown
by the filled symbols in Fig.~\ref{fig:op}. 
On a $26^3$ lattice with $\ls=1$ (with the same physical volume) 
we find that with $b_2=0.084$ and $b_1$ unchanged from above, 
the order parameter as a function of $T$ 
is the same as that for $\ls=0.4$ within error bars 
(open symbols in Fig.~\ref{fig:op}).
We have also verified that the $\ls$-dependence
of $b_2$ vanishes for small $\lambda$  and is
non-divergent in the $\ls\rightarrow 0$ limit.

\begin{figure}[htb]
\psfig{file=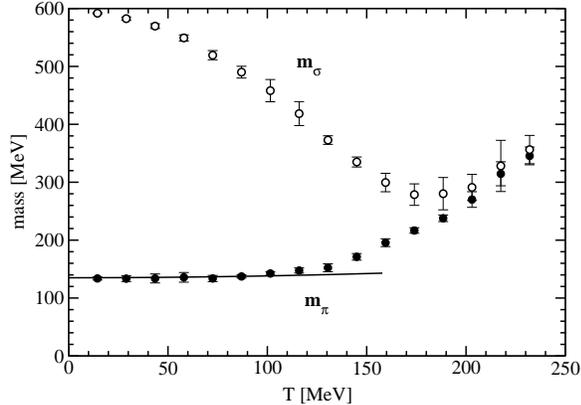,width=2.6in,angle=270}
\vspace{-1 cm}
\caption{Thermal masses of the $\sigma$ and $\pi$ fields 
{\it vs. } T. Error bars denote statistical and best fit uncertainties. 
The line shows the prediction from chiral perturbation theory, 
Eq.~(\ref{mpichiralpt}).
The minimum of $m_\sigma$ is reached at 
$T_{\rm cross}\simeq  180$ MeV. Calculations were done with 
$a_s=0.4$ on a $64^3$ lattice.}
\label{fig:masses}
\vspace{-0.8 cm}
\end{figure}

Next we measured the $T$ dependence of the masses and decay rates 
of the pion and $\sigma$ fields. These quantities 
follow from the dynamical response to small amplitude perturbations
around equilibrium. Then the  response takes the form  
\begin{eqnarray}
\langle\phi_a\rangle 
= \langle\phi_a(t=0)\rangle \exp [- t/\tau] \cos(\omega t), \label{phifit}
\end{eqnarray}
with $\omega = \sqrt{m^2(T) - \tau^{-2}}$, for the particular case 
of a spatially homogeneous perturbation. 
Results are shown in Fig.~\ref{fig:masses}, together with the chiral 
perturbation theory prediction for the pion mass~\cite{chpt}
\begin{equation}\label{mpichiralpt}
m_\pi^2(T)=m_\pi^2(0)\left(1+{T^2 \over 24 f_\pi^2}\right)\ ,
\qquad T\ll f_\pi \ .
\end{equation}

The $\sigma$ becomes lighter in medium as a result of 
symmetry restoration at high $T$. $m_\sigma$ has a minimum
at $T_{\rm cross}\simeq 180$ MeV, which defines the 
crossover temperature. 
At $T_{\rm cross}$, $m_\pi\simeq 220$ MeV and $m_\sigma\simeq 280$ MeV.
The two masses are equal within error bars for $T \simgt 220$ MeV. 

The timescale $\tau(T)$ for the decay of a region of disoriented chiral 
condensate in contact with an equilibrated gas of hadrons has been computed 
previously by Steele and Koch~\cite{SteeleKoch}, using a hadron gas model
and perturbatively (at 2-loops) by Rischke \cite{Rischke}.
We seek to choose $\eta(T)$ so that the $\tau(T)$ we measure 
reproduces their calculations.

\begin{figure}[htb]
\psfig{file=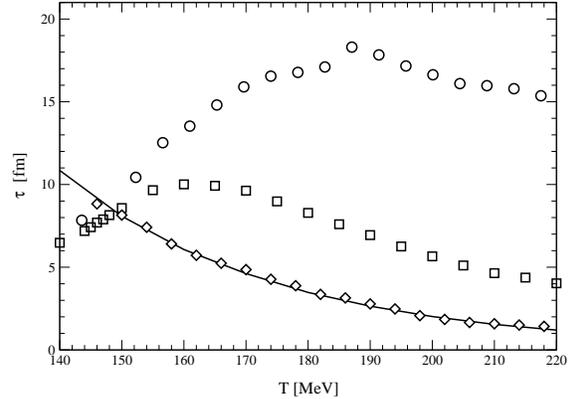,width=2.6in,angle=270}
\vspace{-1 cm}
\caption{Circles show the DCC decay time $\tau_{\rm cl}(T)$ 
for the  classical $\sigma$-model at 
$\eta=0$.   Squares show $\tau(T)$ 
from the Langevin equation with $\eta(T)$ 
taken from Rischke's perturbative calculation.
Diamonds show $\tau(T)$ from the Langevin equation 
with $\eta(T)$ chosen to reproduce $\tau(T)$ 
from the calculation of Steele and Koch (solid line).}
\label{fig:taus}
\vspace{-0.6 cm}
\end{figure}

A difficulty is that the classical thermal field theory already 
leads to long wave length decay. The associated time scale is much 
shorter (larger dissipation) than the target quantum $\tau$ 
at low $T$, because of the absence of Bose-Einstein suppression of 
short wave lengths in the classical case.

Because $\eta \geq 0$  in the Langevin equation (1) 
we find that our model can only be used to describe the long wavelength 
dynamics in the presence of a heat bath with  $T \simgt 145$ MeV,
corresponding to about 80\% of  $T_{\rm cross}$. 
Above this temperature we can adjust our input 
bare $\eta$ to reproduce the DCC decay time computed by Steele and 
Koch, see Fig.~\ref{fig:taus}. 
However, $T \sim 145$ MeV is large enough that the 
assumptions in the calculation of $\tau$ as in 
Ref.~\cite{SteeleKoch} may be starting to break
down, which means that our estimate of the limit of validity of our
analysis may have a little play in it. The same considerations
apply to the perturbative 2-loop calculation, with the additional 
caveat that at $\lambda=20$ it may be unreliable.

In spite of these difficulties we are currently studying the 
non-equilibrium dynamics of the system induced by the combined effect 
of the cooling of the thermal bath and by volume expansion. 
The latter becomes the principal means of cooling at low $T$,
which may help mitigate the problems associated with the choice of $\eta$.

\end{document}